\documentclass[preprint,12pt]{elsarticle}

\usepackage{amssymb}
\usepackage{amsmath} 
\usepackage{color}
\usepackage{hyperref} 
\usepackage{caption}
\usepackage{subcaption}
\usepackage{soul}
\usepackage{lineno}
\usepackage{comment}
\usepackage{relsize}
\usepackage{algorithm}
\usepackage[noend]{algpseudocode}

\makeatletter
\def\BState{\State\hskip-\ALG@thistlm}
\makeatother
\usepackage[symbol]{footmisc}

\begin{document}

%\linenumbers
\begin{frontmatter}

\title{Simplified models for unsteady three-dimensional flows in slowly varying microchannels}

%\cortext[cor1]{Corresponding author Tel: +9611786456, ext 1298, Email: leila.issa@lau.edu.lb}

%\author[lau]{Leila Issa\corref{cor1}}

\author{Leila Issa\footnote[1]{Corresponding author}}
\address{ Department of Computer Science and Mathematics\\
	Lebanese American University\\
	Beirut, Lebanon\\
    Email: leila.issa@lau.edu.lb 	
}

\author{Sajed Medlej}
\address{ Department of Mechanical Engineering\\
	American University of Beirut\\
	Beirut, Lebanon\\
    Email:shm27@mail.aub.edu	
}

\author{Ali Saleh}
\address{ Department of Mechanical Engineering\\
	American University of Beirut\\
	Beirut, Lebanon\\
    Email:ams125@mail.aub.edu
}

\author{Issam Lakkis}
\address{ Department of Mechanical Engineering\\
	American University of Beirut\\
	Beirut, Lebanon\\
    Email:il01@aub.edu.lb    	
}

\begin{abstract}
We present a reduced order model for three dimensional unsteady pressure-driven flows in micro-channels of variable cross-section.
This fast and accurate model is valid for long channels, but allows for large variations in the channel's cross-section along the axis. It is based on an asymptotic expansion of the governing equations in the aspect ratio of the channel. A finite Fourier transform in the plane normal to the flow direction is used to solve for the leading order axial velocity. The corresponding pressure and transverse velocity are obtained via a hybrid analytic-numerical scheme based on recursion. 
The channel geometry is such that one of the transverse velocity components is negligible, and the other component, in the plane of variation of channel height, is obtained from combination of the corresponding momentum equation and the continuity equations, assuming a low degree polynomial Ansatz of the pressure forcing.
A key feature of the model is that it puts no restriction on the time dependence of the pressure forcing, in terms of shape and frequency, as long as the advective component of the inertia term is small. This is a major departure from many previous expositions which assume harmonic forcing. The model reveals to be accurate for a wide range of parameters and is two orders of magnitude faster than conventional three dimensional CFD simulations. 
\end{abstract}


\end{frontmatter}

\section{Introduction}\label{section:intro}
Microfluidic devices allow the operation of fluids inside channels whose typical dimensions are on the order of microns.
Since the micro scale matches the scale of some biological structures, one of the main advantages of these devices is that they make it possible to have multiple biological/chemical functions (a lab) in a small area (a chip).
Applications that use these devices include the transport of biological materials such as proteins, DNA, cells, or transport of chemical samples and analytes. Other applications include sorting, mixing, drug delivery or mass spectroscopy.  For a more detailed exposition of these applications, the reader is referred to the book of \cite{karniadakisbook}.

Of particular importance to these applications is the ability to accurately, yet efficiently predict unsteady flows in in three dimensional channels. Whenever we need to optimize the design of a microfluidic device, we need to swipe many dimensions and forcing signal parameters; this problem becomes prohibitively expensive to solve; specially if one resorts uniquely to using numerical solutions of the full system of the governing P.D.E. (typically CFD tools).
The contribution of this work is to come up with an unsteady three dimensional reduced order model that is fast, yet accurate, to perform these simulations.

%in future direction about an optimization or inverse??} \cite{liu2019nonlinear}.\\
The alternative to fully simulate the governing equations numerically \cite{bhattacharyya2015combined} is to resort to reduced order models. These can be based on simplifying the governing equations based on scaling. This is where our work fits. 

%This model is perturbation based and semi-analytic unsteady models These models are valid in the low Reynolds number and long micro channels regime.
Our model is based on approximating the Navier-Stokes equations using a regular perturbation expansion in the aspect ratio of the channel. It turns out that the resulting systems of equations can be partially solved for the leading order axial velocity using projections onto a space of appropriate eigenfunctions. The use of a recursive method to fully solve for the time dependence of the pressure distribution, and the fact that these expressions need only a few number of modes to converge render these models much faster than CFD. To solve for the transverse velocity components, $v$ and $w$, we assume that the channel geometry is such that $w$ is negligible; The $v$ component 
is obtained from the corresponding momentum and continuity equations, assuming a low degree polynomial ansatz of the pressure forcing. 
%With this approach, solving for $v$ requires numerically solving the resulting linear equation governing the polynomial coefficients. One of the advantages of this approach is decoupling $u$ from $v$ and $w$. In this manner, one can opt for an approximate solution for $u$ and $p$ only, or additionally, choose to solve for $v$ at selected cross sections normal to the flow direction.}

%\cite{GAMRAT20052943}. %Some studies concerning flows between parallel plates also account for geometry irregularities such as  
%solutions are numerical or hybrid if ...

%Unsteady flows are much less investigated, even in parallel channels \cite{Fan65}.
When the geometry is simple and the flow is steady, analytical solutions are possible. For example, this is the case for steady, laminar and fully developed flows in channels of constant cross section with regular geometry, such as rectangular or circular cross sections. 
For more complex geometries, solutions are often numerical or hybrid. In particular, some works focus on flows in channels of constant cross section with irregular geometry \cite{pinheiro90integral}. 
%A hybrid numerical-analytical solution, based on the Generalized Integral Transform Technique, is possible and the applications are to a trapezoid-like duct to account for fabrication imperfections, as commonly found in commercially manufactured microchannels. 
Flows in narrow and long geometries whose cross section is not constant are frequently investigated using the theory of lubrication. This theory was used to approximate the steady Navier-Stokes and continuity equations in the low Reynolds number regime, for slowly variable geometries 
\cite{batchelor} and later adapted to more general contexts.
 There is a vast literature on lubrication based models in the steady and two dimensional case, and those typically focus on obtaining to leading order, mass-flow relations. For a more detailed account on those, see e.g. the introduction in \cite{issa2016perturbation,ISSA2018410}.
 As for flows in three dimensional rectangular channels, the work in \cite{lauga2004three} focuses on the steady case. The authors there use lubrication theory to show that if the cross-section is varying in the flow-wise direction, then the flow cannot be planar. 
 
Concerning research on unsteady flows, the assumption is often that the pressure forcing is harmonic; for example this was done in the context of modeling bio-flows (see for e.g. \cite{rao1973pulsatile,hall1974unsteady}). Before we proceed, we clarify here that what we mean by unsteady is the unsteadiness of the flow manifested in the inertia term, i.e., the time varying component of inertia, which we assume is non negligible.  
For instance the authors of \cite{hall1974unsteady} analyzed a pressure driven oscillatory flow in a circular channel of slowly varying cross-section, allowing small departure from symmetry. Their solution is developed for a harmonically oscillatory forcing with the oscillation frequency being too small $(\mathtt{ReSt} << 1)$ or very large $(\mathtt{ReSt} >> 1)$.

 In the context of lubrication theory, and up to our knowledge, this would be the first work to focus on three dimensional and unsteady flows in variable channels, with non-harmonic pressure forcing. The model we are proposing falls within the scope of the work done in \cite{issa2016perturbation,ISSA2018410,issa2013reduced,issa2014reduced,lakkis2008}.
 In particular, 
it is an extension of the unsteady model developed for two dimensional rectangular channels in \cite{issa2016perturbation} to three dimensions, while relaxing the nearly parallel channel assumption. 
 A key contribution of this work is that the model puts no restriction on the time dependence of the forcing, in terms of shape and frequency. The only condition on the forcing is such that the advective component of the inertia term is small. This is a major departure from these previous expositions that assume harmonic forcing. 
Another important feature of these models is that they allow us to extract ranges of validity in terms of the dependence of the error distributions of velocities and pressure on the parameters of interest. 

This paper is organized as follows: in sections \ref{ss:nondim} and \ref{ss:BoundaryConditions}, we present the governing equations and the boundary conditions with the relevant scaling, and then, in section \ref{ss:leadingEq}, we obtain the leading order equations from them. Next, in section \ref{sss:u0p0},  we present the semi-analytical solutions for $u$ and $p$ to these equations using a finite Fourier transform in the plane normal to the flow direction. Our approach to solving for the leading lateral and transverse velocities is presented in section \ref{sss:vw}. In the numerical results section \ref{section:results}, we show dependence of the error on key parameters for selected cases, in addition to comparisons of the pressure profiles and velocity contours obtained from our model, with those obtained from ANSYS. In section {section:Conclusion}, we conclude and propose future directions.

%\textcolor{red}{We then present our approach to solve for the transverse velocity components, $v$ and $w$, in section AGOOGI we assume that the channel geometry is such that $w$ is negligible, and the other component, $v$ in the plane of variation of channel height (see Figure ), is obtained from combination of the corresponding momentum equation and the continuity equation, assuming a low degree polynomial anzats of the pressure forcing. With this approach, solving for $v$ requires numerically solving the resulting linear equation governing the polynomial coefficients. One of the advantages of this approach is decoupling $u$ from $v$ and $w$. In this manner, one can opt for an approximate solution for $u$ and $p$ only, or additionally, choose to solve for $v$ at selected cross sections normal to the flow direction.}

%\textcolor{red}{anything about v here?} 

%\label{section:intro}

%\label{section:model}
%	\label{ss:nondim} 
%	\label{ss:BoundaryConditions}
%	\label{ss:leadingEq}
%		\label{sss:u0p0}
%		\label{sss:vw}

%\label{section:results}

%\label{section:Conclusion}

%%%%%%%%%%%%%%%%%%%%%%%%%%%%%%%%%%%%%%%%%%%%%%%%%%%%
%%%%%%%%%%%%%%%%%CLEAN PAPER VERSION%%%%%%%%%%%%%%%%%%%%%%%%%%%%%%%%%%
\section{Model of an unsteady flow in three dimensional shallow channels}\label{section:model}

As illustrated in Fig. 1, we consider a pressure driven flow along  the $\hat x$ direction, in a three dimensional micro-channel with constant height $z=2H_0,$ and a slowly varying cross-section $y=D(x)$. The typical axial length scale over which the variations in the cross-section occur define the length scale $L_x$. We will assume in this work that the aspect ratio defined as $\alpha = \frac{H_0}{L_x}$ is small, i.e. $\alpha << 1$. This will allow us later to formulate a  solution based on an asymptotic expansion of our fields in the parameter $\alpha$.

%The channel is symmetric in the $\hat y$ and $\hat z$ directions. This assumption results in Neumann boundary conditions at the center-line and Dirichlet's at the edges.
\subsection{Governing equations in non-dimensional form}\label{ss:nondim} 
The continuity equation and momentum conservation for an unsteady incompressible flow are
\begin{equation}
\begin{aligned}
\partial_x u + \partial_y v + \partial w_z & =0, \\
\rho \Big( \partial_t u + \textbf u . \nabla u \Big) & = -\partial_x p + \mu \Delta u , \\
\rho \Big( \partial_t v + \textbf u . \nabla v \Big) & = -\partial_y p + \mu \Delta v , \\
\rho \Big( \partial_t w + \textbf u . \nabla w \Big) & = -\partial_z p + \mu \Delta w , \\
\end{aligned}
\label{eq:goveqns}
\end{equation}

The gravitational body forces are neglected in the applications of micro-flows with dominant pressure variation $p(x,y,z,t)$. The fluid is considered incompressible with a constant density $\rho$, a velocity vector $\textbf u = (u, v, w)$, and constant fluid dynamic viscosity $\mu$.

\begin{figure}[!htp]
\centering
\includegraphics[width=5in]{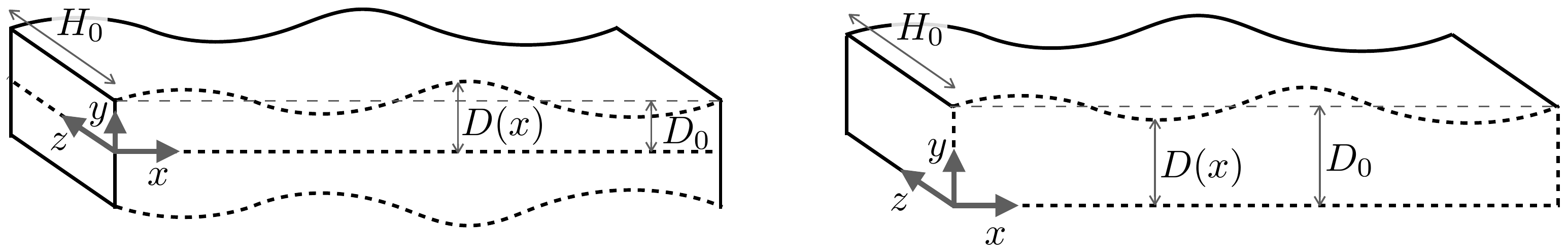}
\caption{Left:  channel Type I. Surfaces $y=D(x)$ and $z=H_0$ are solid impermeable walls. Planes $y=0$ and $z=0$ are planes of symmetry. Aspect ratio $A = \frac{D_0}{H_0}$.
Right:  channel Type II. Surfaces $y=0$, $y=D(x)$ and $z=H_0$ are solid impermeable walls. Plane $z=0$ is a plane of symmetry. Aspect ratio $A = \frac{D_0}{2H_0}$.}
\label{fig:channelgeo}
\end{figure}

The equations in ~\ref{eq:goveqns} are nondimensionalized in order to show the different characteristic parameters of the flow, which allows the proper scaling of the physical terms based on the the geometric and physical assumptions of the problem. Hence, the following dimensionless parameters are used, with a star coupled to each
\begin{gather*}
x^{\star}= \frac{x}{L_x}, z^{\star}= \frac{z}{H_0}, y^{\star}= \frac{y}{D_0}, t^{\star}= tf, \\ u^{\star}= \frac{u}{U}, v^{\star}= \frac{v}{\alpha U}, w^{\star}= \frac{w}{\alpha^2 U}, p^{\star}= \frac{p}{\Delta P_0}
\end{gather*}

Since the flow in the channel is pressure-driven, the pressure force in $\hat x$-direction per unit volume can be scaled as $\frac{\Delta P_0}{L_x}$, where $\Delta P_0$ is the pressure drop corresponding to a Poiseuille flow. The following Reynolds, Strouhal, and Poiseuille dimensionless numbers are also introduced into the governing equations
\[
\mathtt{Re} \equiv \frac{\rho U H_0}{\mu}, \thinspace \mathtt{St} \equiv \frac{fH_0}{U}, \thinspace \mathtt{Po} \equiv \frac{\Delta P_0 / L_x}{\mu U/H_0^2}.
\]

The continuity and momentum equations in \ref{eq:goveqns} become in dimensionless form after dropping the stars and defining $A=D_0/H_0$
\begin{align} 
&\partial_x u +\frac{1}{A} \partial_y v + \partial_z w  =0, \\ \nonumber
&\mathtt{Re}\Big( \mathtt{St}\partial_t u+ \alpha\Big( u\partial_x u+\frac{v}{A}\partial_y u+ w\partial_z u\Big) \Big) \\  \nonumber
&= -\mathtt{Po} \partial_x p + \alpha^2\partial^2_{x^2} u +\frac{1}{A^2}\partial^2_{y^2} u + \partial^2_{z^2} u, \\   \nonumber
&\mathtt{Re}\alpha^2\Big( \mathtt{St}\partial_t v+ \alpha \Big( u\partial_x v + \frac{v}{A}\partial_y v+ w\partial_z v\Big) \Big)\\  \nonumber
 &= -\mathtt{Po} \partial_y p +\alpha^2 \Big( \alpha^2 \partial^2_{x^2} v +\frac{1}{A^2}\partial^2_{y^2} v+ \partial^2_{z^2} v\Big), \\   \nonumber
 &\mathtt{Re}{\alpha^3}\Big( \mathtt{St}\partial_t w + \alpha \Big( u\partial_x w+\frac{v}{A}\partial_y w+ w\partial_z w\Big) \Big)\  \\ \nonumber
  &= -\mathtt{Po} \partial_z p +\alpha^2\Big( \alpha^2 \partial^2_{x^2} w +\frac{1}{A^2}\partial^2_{y^2} w +\partial^2_{z^2} w\Big).  
\end{align}

%\begin{align} 
%&\frac{\partial u}{\partial x} +\frac{1}{A} \frac{\partial v}{\partial y} +\frac{\partial w}{\partial z}  =0, \\ \nonumber
%&\mathtt{Re}\Big( St\frac{\partial u}{\partial t}+ \alpha\Big( u\frac{\partial u}{\partial x}+\frac{v}{A}\frac{\partial u}{\partial y}+ w\frac{\partial u}{\partial z}\Big) \Big) \\  \nonumber
%&= -\mathtt{Po} \frac{\partial p}{\partial x} + \alpha^2\frac{\partial^2 u}{\partial x^2}+\frac{1}{A^2}\frac{\partial^2 u}{\partial y^2}+\frac{\partial^2 u}{\partial z^2}, \\   \nonumber
%&\mathtt{Re}\alpha^2\Big( St\frac{\partial v}{\partial t}+ \alpha \Big( u\frac{\partial v}{\partial x}+\frac{v}{A}\frac{\partial v}{\partial y}+ w\frac{\partial v}{\partial z}\Big) \Big)\\  \nonumber
% &= -\mathtt{Po} \frac{\partial p}{\partial y} +\alpha^2 \Big( \alpha^2\frac{\partial^2 v}{\partial x^2}+\frac{1}{A^2}\frac{\partial^2 v}{\partial y^2}+\frac{\partial^2 v}{\partial z^2}\Big), \\   \nonumber
% &\mathtt{Re}\textcolor{red}{\alpha^3}\Big( St\frac{\partial w}{\partial t}+ \alpha \Big( u\frac{\partial w}{\partial x}+\frac{v}{A}\frac{\partial w}{\partial y}+ w\frac{\partial w}{\partial z}\Big) \Big)\  \\ \nonumber
%  &= -\mathtt{Po} \frac{\partial p}{\partial z} +\alpha^2\Big( \alpha^2\frac{\partial^2 w}{\partial x^2}+\frac{1}{A^2}\frac{\partial^2 w}{\partial y^2}+\frac{\partial^2 w}{\partial z^2}\Big)  
%\end{align}

\subsection{Boundary Conditions}\label{ss:BoundaryConditions}
Two types of boundary conditions are considered

\begin{enumerate}

\item Type I (the plane $y=0$ is a symmetry condition)
\begin{equation}
\begin{aligned}
\partial_y u(x,0,z,t)&=0=\partial_y w(x,0,z,t) &\text{and} \thinspace\thinspace\thinspace v(x,0,z,t)=0, \,\,t>0
\\ \notag
\partial_z u(x,y,0,t)&=0=\partial_z v(x,y,0,t) &\text{and} \thinspace\thinspace\thinspace w(x,y,0,t)=0, \,\, t>0
\end{aligned}
\label{eq:bcaxisym2}
\end{equation}

\begin{equation}
\begin{aligned}
u(x,D(x),z,t)&=v(x,D(x),z,t)=w(x,D(x),z,t)=0, \,\, t>0 \notag
\\u(x,y,1,t)&=v(x,y,1,t)=w(x,y,1,t)=0, \,\,t>0
\end{aligned}
\label{eq:bcnoslip2}
\end{equation}

\item Type II (the plane $y=0$ is a wall)
\begin{equation}
\begin{aligned}
u(x,0,z,t)&=w(x,0,z,t)=v(x,0,z,t)=0,\,\, t>0
\\ \notag
\partial_z u(x,y,0,t)&=\partial_z v(x,y,0,t)=0, \text{and} \thinspace\thinspace\thinspace w(x,y,0,t)=0, \,\,t>0
\end{aligned}
\label{eq:bcaxisym}
\end{equation}

\begin{equation}
\begin{aligned}
u(x,D(x),z,t)&=v(x,D(x),z,t)=w(x,D(x),z,t)=0, \,\,t>0 \notag
\\u(x,y,1,t)&=v(x,y,1,t)=w(x,y,1,t)=0, \,\, t>0
\end{aligned}
\label{eq:bcnoslip}
\end{equation}

\end{enumerate}

where $D(x)$ is the nondimensional channel depth $D^{\star}(x)=D(x)/D_0$ after removing the star, and initial conditions are
\begin{equation}
\begin{aligned}
u(x,y,z,0)=v(x,y,z,0)&=w(x,y,z,0)=0
\\p(x,y,z,0)&=p_a
\end{aligned}
\label{eq:ic}
\end{equation}

\subsection{Leading Order Equations}\label{ss:leadingEq}
The relevant fields are expanded asymptotically in the aspect ratio $\alpha$ which is assumed to be small ($\alpha << 1$)
\begin{equation}
\phi=\phi_0 +\alpha \thinspace \phi_1 + \cdots,
\label{eq:expansion}
\end{equation}
where $\phi$ could be $u,v,w$ or $p$. Since the pressure boundary conditions at the entrance and exit of the channel are essential ($p_0$ is known there), the higher-orders of pressure vanish at these two locations. This implies that the initial condition for the zeroth order pressure distribution is $p_0(x,y,z,0)=p_a$.
%In the following derivations, all other parameters including $ Re$ and $St$ are considered of order of unity for simplicity. The solution still holds for small $Re$ (i.e. $Re<1$) and for all values of $St$.

The asymptotic expansion yields the zeroth order equations ($\sim 1$ ), with velocities inheriting the same boundary conditions at the centerline and walls as in \ref{ss:BoundaryConditions}

\begin{align}\label{eq:zeroth}
\partial_x u_0 &+ \frac{1}{A}\partial_y v_0 +\partial_z w_0 =0, 
\\ \mathtt{Re St}\partial_t u_0&=-\mathtt{Po} \partial_x p_0+\frac{1}{A^2}\partial^2_{y^2} u_0+\partial^2_{z^2} u_0, \nonumber \\ 
0&=-\mathtt{Po} \partial_y p_0, \nonumber \\  0&=-\mathtt{Po} \partial_z p_0 \nonumber.
\end{align}
In what follows, and for the sake of simplicity of exposition, we only present the solutions for the boundary conditions of Type I. Type II conditions can be dealt with in a very similar way. 

%and the first-order equations ($\sim \alpha$) are
%\begin{align}
%&\frac{\partial u_1}{\partial x} + \frac{1}{A}\frac{\partial v_1}{\partial y} +\frac{\partial w_1}{\partial z} =0,
%\\  & \mathtt{Re St}\frac{\partial u_1}{\partial t}+\mathtt{Re} %\Big( u_0\frac{\partial u_0}{\partial %x}+\frac{1}{A}v_0\frac{\partial u_0}{\partial y}+ w_0\frac{\partial u_0}{\partial z}\Big)=-\mathtt{Po} \frac{\partial p_1}{\partial x}+ \nonumber \\
%&\frac{1}{A^2}\frac{\partial^2 u_1}{\partial y^2}+\frac{\partial^2 u_1}{\partial z^2}, \nonumber \\  0&=-\mathtt{Po} \frac{\partial p_1}{\partial y},  \nonumber \\  0&=-\mathtt{Po} \frac{\partial p_1}{\partial z} \nonumber
%\end{align}

%%%%%%%%%%%%%%%%%%%%%%%%%%%%%%%%%%%%%%%%%%%%%%%%%%%%%%%%%%%%%%%%%%%%%%%%%%%%%%%%%%%%%%%%%%%%%%%%%%%%%%%%%%%%%%%%%%%%%%%%%%%%%%%%%%%%%%%%%%%%%%%%%%%%%%%%%%%%%%%%%%%%%%%%%%%%%%%%%%%%%%%%%%%%%%%%%%%

\subsection{Leading order velocities and pressure for Type I channels}

\subsubsection{Axial velocity and pressure}\label{sss:u0p0}
The leading order axial velocity is obtained using a finite Fourier transform in the plane orthogonal to the flow direction on the momentum equation
\begin{align}
&u_0 (x,y,z,t)=\notag \\
&\frac{-4\mathtt{Po}}{\mathtt{Re}\mathtt{St}}\sum_{m,n=1}^{\infty} \frac{(-1)^{m+n}}{\nu_m \nu_n} \cos{\Big(\nu_m \frac{y}{D(x)}\Big)} \cos{(\nu_n z)} I_{m,n}\bigg( \frac{\partial p_0}{\partial x}(x,t)\bigg),
\end{align}
where $\nu_n=\pi(n-1/2), \,\, n=1,2,\cdots,$ and where we have introduced the following quantities for compactness
\begin{gather*} 
I_{m,n}\big(f(x,t)\big)=\int_{t'=0}^t e^{-\kappa_{m,n}(t-t')}f(x,t')dt', \\
\kappa_{m,n}(x)=\frac{\frac{\nu_m^2}{A^2\, D^2(x)}+\nu_n^2}{\mathtt{Re}\mathtt{St}}. 
\end{gather*}

This expression depends on the unknown pressure $p_0,$ which is obtained implicitly from the following methodology:
we first integrate the continuity equation along the cross-sectional area normal to $\hat x$, and realize that the volumetric flow rate in the channel is independent of the flow direction, i.e. 
\begin{equation}
Q_0(t)=\int_{-1}^{1} \int_{-D(x)}^{D(x)} u_0 \, dx \, dy=  - \frac{16 \mathtt{Po}}{\mathtt{Re}\mathtt{St}}\sum_{m,n=1}^{\infty}\frac{D(x)}{\nu_m^2\nu_n^2}I_{m,n}\big( \partial_x p_0(x,t) \big).
\label{eq:Q0TYPE1}
\end{equation}

Equation (\ref{eq:Q0TYPE1}) is then solved recursively in time with
\[
Q_0(t+\Delta t)=-\frac{16\mathtt{Po}}{\mathtt{Re}\mathtt{St}}\sum_{m,n=1}^{\infty}\frac{D(x)}{\nu_m^2\nu_n^2}I_{m,n}\big( \partial_x p_0(x,t+\Delta t) \big)
\]

where

\begin{align}
&I_{m,n}\big( \partial_x p_0(x,t+\Delta t) \big)=e^{-\kappa_{m,n}\Delta t}\cdot I_{m,n}\big( \partial_x p_0(x,t)\big) \notag \\
&+ \int_t^{t+\Delta t}e^{-\kappa_{m,n}(t+\Delta t -t')}\partial_x p_0(x,t')dt'.
\label{eq:Imnplusdeltat}
\end{align}

To resolve the last term in this equation, a quadrature with coefficients $a_{1,2}^{m,n}$ is used, and the integral reduces to
\begin{align}
&\int_t^{t+\Delta t}e^{-\kappa_{m,n}(t+\Delta t -t')}\partial_x p_0(x,t')dt' \\ \notag
&=a_1^{m,n}(x,\Delta t)\partial_x p_0(x,t)+a_2^{m,n}(x,\Delta t)\partial_x p_0(x,t+\Delta t).
\label{eq:ImnQUAD}
\end{align}

This yields the flow rate at $t+\Delta t$
 
\begin{align}
Q_0(t+\Delta t)&=-\frac{16\mathtt{Po}}{\mathtt{Re}\mathtt{St}}\sum_{m,n=1}^{\infty}\frac{D(x)}{\nu_m^2\nu_n^2}\Big[ e^{-\kappa_{m,n}\Delta t}I_{m,n}\big( \partial_x p_0(x,t)\big) \notag \\
&+ a_1^{m,n}(x,\Delta t)\partial_x p_0(x,t)+a_2^{m,n}(x,\Delta t)\partial_x p_0(x,t+\Delta t) \Big] \notag \\
&=-4C_0 D(x) \bigg( \sum_{m,n=1}^{\infty}\frac{1}{\nu_m^2\nu_n^2} e^{-\kappa_{m,n}\Delta t}I_{m,n}\big( \partial_x p_0(x,t)\big) \notag \\
&+ \partial_x p_0(x,t) S_1(x) \bigg.+ \bigg. \partial_x p_0(x,t+\Delta t) S_2(x) \bigg).
\end{align}

Here, we have defined the following parameters for compactness
\begin{gather*}
C_0=\frac{4\mathtt{Po}}{\mathtt{Re}\mathtt{St}} \\
S_1(x)=\sum_{m,n=1}^{\infty}\frac{a_1^{m,n}(x,\Delta t)}{\nu_m^2\nu_n^2} \\
S_2(x)=\sum_{m,n=1}^{\infty}\frac{a_2^{m,n}(x,\Delta t)}{\nu_m^2\nu_n^2}.
\end{gather*}

%\begin{itemize}
%\item $C_0=\frac{4\mathtt{Po}}{\mathtt{Re}\mathtt{St}}$
%\item $S_1(x)=\sum_{m,n=1}^{\infty}\frac{a_1^{m,n}(x,\Delta t)}{\nu_m^2\nu_n^2}$ 
%\item $S_2(x)=\sum_{m,n=1}^{\infty}\frac{a_2^{m,n}(x,\Delta t)}{\nu_m^2\nu_n^2}.$
%\end{itemize}

Upon integrating in $x$ from inlet to outlet, and after some algebraic manipulations, we obtain the volume flow rate at $t+\Delta t$ in terms of the pressure drop $\Delta P_0$ as

\begin{align}\label{eq:dQ0iterI}
&Q_0(t+\Delta t) =   \frac{{4} C_0}{  \int_0^{L/L_x} \frac{1}{D(x) S_2(x)} \, dx }  \bigg[ \Delta P_0(t+\Delta t)
 \notag \\
&-\int_0^{L/L_x} \frac{1}{S_2(x)}\sum_{m,n=1}^{\infty}\frac{1}{\nu_m^2\nu_n^2} e^{-\kappa_{m,n}\Delta t}I_{m,n}\big(\partial_x p_0(x,t)\big) \, dx \notag \\
&-  \int_0^{L/L_x} \frac{S_1(x)}{S_2(x)} \partial_x p_0 (x,t)\, dx \bigg].
\end{align}

This allows us to use the following relation to update the pressure drop
\begin{align}\label{eq:dp0iterI}
&\partial_x p_0(x,t+\Delta t)= - \frac{1}{4C_0 D(x) S_2(x)} Q_0(t+\Delta t) 
 \notag \\
&- \frac{1}{S_2(x)}\sum_{m,n=1}^{\infty}\frac{1}{\nu_m^2\nu_n^2} e^{-\kappa_{m,n}\Delta t}I_{m,n}\big( \partial_x p_0(x,t)\big)- \frac{S_1(x)}{S_2(x)} \partial_x p_0 (x,t)
\end{align}

We summarize the procedure that allowed us to obtain the pressure in Algorithm 1.
\begin{algorithm}[!h]
\caption{Algorithm for computing $p_0$}\label{euclid}
\begin{algorithmic}[1]
%\Procedure{MyProcedure}{}
%\State hi
%\State hi
\BState \emph{precompute}: \begin{itemize}
\item number of modes $N$ 
\item time independent quantities:  $S_1(x)=\sum_{m,n=1}^N\frac{a_1^{m,n}}{\nu_m^2\nu_n^2}$ and $S_2(x)=\sum_{m,n=1}^N\frac{a_2^{m,n}}{\nu_m^2\nu_n^2}$.
 \end{itemize}
%\If {$i > \textit{stringlen}$} \Return false
%\EndIf
%\State hi
\BState \emph{ initialize $\frac{\partial p_0}{\partial x}(x,t=0)=0$ and $I_{m,n}\Big( \frac{\partial p_0}{\partial x}(x,t=0) \Big)=0$} 
\BState \emph{loop over time}: $t \rightarrow t+\Delta t$, compute
\begin{itemize}
	\item Compute $Q(t+\Delta t)$ using the recurrence relation (\ref{eq:dQ0iterI})
	\item Update $\frac{\partial p_0}{\partial x}(x,t+\Delta t)$ using the recurrence relation (\ref{eq:dp0iterI})
	\end{itemize}%\EndProcedure
\end{algorithmic}
\end{algorithm}

%%%%%%%%HERE%%%%%%%%%%

%\input

\subsubsection{Transverse and lateral velocities}\label{sss:vw}
To leading order, the $w$ component of velocity is zero according to our scaling. So one could use the conservation of mass to obtain $v_0$ from $u_0$. However, the trouble with this approach is that the solution obtained is not consistent with the boundary condition at the wall. In what follows we present a novel method to obtain a transverse velocity that satisfies the desired boundary conditions. 

%$w_0 \sim 0$ If one were to use the continuity equation to obtain $v_0$
%\textcolor{red}{Explain why we can't simply use the continuity equation to get v}
%We now assume that $W/U$ scales as $\alpha^2.$, the equation of $u_0$ is unchanged, $w_0\sim 0$ to this order, and we can get $v_0$ from the continuity equation
%\begin{align}
%v(x,y,z,t) = - \int \frac{\partial u}{\partial x} dy + g(x,z,t)
%\end{align}
%where $v$ is subject  to the boundary conditions presented in Section \ref{ss:BoundaryConditions}. Enforcing the boundary condition $v(y=0)=0$ yields $g(x,z,t) = \left( \int \frac{\partial u}{\partial x} dy \right)_{y=0} = 0$ so that
%\begin{align}
%v(x,y,z,t) = - \int \frac{\partial u}{\partial x} dy 
%\end{align}
%It can be shown that this solution satisfies the boundary conditions $(\partial v/\partial z)=0$ at $z=0$ and $v=0$ at $z=1$. At $y=D(x)$, 
%
%\begin{align} 
%&v_0 (x,y=D(x),z,t) \\ \notag 
%&= \frac{4 A\mathtt{Po}}{\mathtt{Re}\mathtt{St}}
%\sum_{m=0}^{\infty}\sum_{n=0}^{\infty}
% \frac{(-1)^{n}}{\nu_m^2 \nu_n}  
%  \cos{(\nu_n z)} \left( D'(x) I_{m,n}\bigg( \frac{\partial p_0}{\partial x}(x,t) \bigg)  + D(x) \frac{\partial}{\partial x}I_{m,n}\bigg( \frac{\partial p_0}{\partial x}(x,t) \bigg) \right)
%\end{align}
%which is not equal to 0 for $0 \leq z \leq 1$. In what follows, we present  a method to solve for $v$.

According to the $O(\alpha^2)$ equations, the zeroth order equation for $v$ satisfies
\begin{equation}
\mathtt{Re St}  \partial_t v_0 =
-\mathtt{Po} \partial_y p_2+\frac{1}{A^2}\partial^2_{y^2} v_0+\partial^2_{z^2} v_0,
\label{eq:alpha2v0}
\end{equation}
and is forced by the unknown pressure term $\partial_y p_2$. Here, instead of running into the closure problem for the pressure term, we assume that it is for now an unknown function denoted by $f_v(x,y,z,t)$, thus \ref{eq:alpha2v0} yields:
\begin{equation}
\mathtt{Re St}  \partial_t v_0 =
f_v(x,y,z,t)+\frac{1}{A^2}\partial^2_{y^2} v_0+\partial^2_{z^2} v_0,
\label{eq:Vo}
\end{equation}
subject  to the boundary conditions presented in Section \ref{ss:BoundaryConditions}. For the sake of compactness, we present this solution for Type I channel only, but a similar procedure can be reached for Type II channels.

To proceed with a solution methodology similar to that used for $u_0$ in \ref{sss:u0p0}, we assume a polynomial expansion of $f_v(x,y,z,t)$ in the plane orthogonal to the flow direction with coefficients dependent on time and flow direction
\begin{align}
f_v(x,y,z,t) \simeq  \sum_{i=0}^{N_i} \sum_{j=0}^{N_j} C_{ij}(x,t) \eta^{i} z^{j}, 
\label{eq:form}
\end{align}
where $\eta=\frac{y}{A\,D(x)}$.

The solution to (\ref{eq:Vo}) is then given by 

%\begin{align}
%&v(x,y,z,t)=\frac{1}{\mathtt{Re}\mathtt{St}} \frac{4}{ A D(x)} \notag \\
%&\sum_{m,n=1}^{\infty} \sin\left(\frac{m \pi y}{A D(x)} \right) \cos\left(\frac{n_1 \pi z}{2 }\right) \int_0^t e^{-\frac{\left(\frac{m \pi}{A D(x)}\right)^2  + \left(\frac{n_1 \pi}{2 }\right)^2}{\mathtt{Re St}}(t-t')} \int_0^{1} \int_0^{AD(x)} \sin\left(\frac{m \pi y'}{A D(x)} \right) \cos\left(\frac{n_1 \pi z'}{2}\right)  f_v(x,y',z',t') dy' dz' dt' .
%\end{align}

%$\nu_{n_1} = \frac{n_1 \pi}{2}$
\begin{align}
v_0(x,y,z,t)&=\frac{4}{\mathtt{Re}\mathtt{St}} \sum_{m,n=1}^{\infty} \sin\left(\beta_m \eta \right) \cos\left(\nu_{n} z \right) \cdot \\ \notag &\int_0^t e^{-k_{m,n}^{(v)} (t-t')} \int_0^{1} \int_0^{1} \sin\left(\beta_m \eta' \right) \cos\left(\nu_{n} z'\right)  f_v(x,\eta',z',t') d\eta' dz' dt' ,
\end{align}
where the following parameters are introduced
\begin{gather*}
\beta_m = m \pi \\
k_{m,n}^{(v)} =  \frac{\left(\frac{\beta_m}{A D(x)}\right)^2  + \left(\nu_{n}\right)^2}{\mathtt{Re St}}.
\end{gather*}

Substituting the form (\ref{eq:form}) in the expressions for $v_0$ allows to carry out the integrals in $y$ and $z$

%\begin{align}
%&v(x,y,z,t)=\frac{4}{\mathtt{Re}\mathtt{St}} \sum_{m,n=1}^{\infty} \sin\left(\beta_m \eta \right) \cos\left(\nu_{n} z \right)\int_0^t e^{-k_{m,n}^{(v)} (t-t')} \int_0^{1} \int_0^{1} \sin\left(\beta_m \eta' \right) \cos\left(\nu_{n} z'\right)   \sum_{i,j}   \left(C_{ij}(x,t) \eta'^{i}z'^{j} \right)  d\eta' dz' dt' 
%\end{align}
%Expressed as 
\begin{align}
v_0(x,y,z,t)&=\frac{4}{\mathtt{Re}\mathtt{St}} \sum_{m,n=1}^{\infty} \sin\left(\beta_m \eta \right) \cos\left(\nu_{n} z \right) \cdot \\ \notag &\sum_{i,j}  I_{m,n}^{(v)}(C_{ij}(x,t)) \int_0^{1} \int_0^{1} \sin\left(\beta_m \eta' \right) \cos\left(\nu_{n} z'\right)     \eta'^{i}z'^{j}  d\eta' dz' ,
\end{align}
where 
\begin{small}
\begin{align}
I_{m,n}^{(v)}(C_{ij}(x,t))  & = \int_0^t e^{-k_{m,n}^{(v)}(t-t')}  C_{ij}(x, t') dt'
\end{align}
\end{small}
From this, we obtain
\begin{align}
&v(x,y,z,t)=\\ \notag
&\frac{4}{\mathtt{Re}\mathtt{St} } \sum_{m,n=1}^{\infty} \sin\left(\beta_m \eta \right) \cos\left(\nu_{n} z \right) \sum_{i,j}  I_{m,n}^{(v)}(C_{ij}(x,t))  M_S(i,2m) M_C\left(j,2n-1 \right)
\end{align}
where
\begin{small}
\begin{align}
M_C(j,n) & =\int_0^1 z^j \cos\left(\frac{n \pi z}{2} \right)  dz \\
M_S(i,m) & =\int_0^1 y^i \sin\left(\frac{m \pi y}{2} \right)  dy 
\end{align}
\end{small}

We now use the continuity equations to solve recursively for the unknown coefficients $C_{i,j}(x,t)$ as we did for the pressure term in $u_0$. In particular, for $ I_{m,n}^{(v)}(C_{ij}(x,t))  $, we use similar expressions to that in equations (\ref{eq:Imnplusdeltat}) and (\ref{eq:ImnQUAD}), we get that the coefficients $C_{ij}$ are updated in time according to

\begin{footnotesize}
\begin{align}
& \sum_{i,j} C_{ij}(x,t+\Delta t)  \sum_{m,n=1}^{\infty} b_{m,n_1}^{(v)}(x,\Delta t)\frac{\beta_m}{D(x)} \cos\left(\beta_m \eta \right) \cos\left(\nu_{n} z \right) M_S(i,2m) M_C\left(j,2n-1 \right)  \notag \\
&=\mathtt{Po} \sum_{m,n=1}^{\infty} \frac{(-1)^{m+n}}{\nu_{n}}
\frac{D'(x)}{D(x)} \eta \sin{\left(\nu_{m_1} \eta\right)} 
\cos{ \left(\nu_{n} z \right)} I_{m_1,n_1}^{(u)}\left( \partial_x p_0(x,t+\Delta t)\right)\notag \\
&+\mathtt{Po} \sum_{m,n=1}^{\infty} \frac{(-1)^{m+n}}{\nu_{m_1} \nu_{n}} \cos{\left(\nu_{m_1} \eta\right)} \cos{ \left(\nu_{n} z \right)} \frac{\partial }{\partial x} \left\{I_{m_1,n_1}^{(u)}   \left( \partial_x p_0(x,t+\Delta t)\right)\right\} \notag \\
& -\sum_{m,n=1}^{\infty} \frac{\beta_m}{D(x)} \cos\left(\beta_m \eta \right) \cos\left(\nu_{n} z \right) \sum_{i,j} [e^{-\kappa_{m,n_1}^{(v)}\Delta t}\cdot I_{m,n}^{(v)}\left( C_{ij}(x,t)\right)\notag \\
&+ a_{m,n_1}^{(v) }(x,\Delta t)C_{ij}(x,t)]  M_S(i,2m) M_C\left(j,2n-1 \right)  
\label{eqxcvb} 
\end{align}
\end{footnotesize}
We choose to solve for $C_{ij}$ at a given $x$ and $t$ by satisfying equation (\ref{eqxcvb}) at uniformly spaced locations in the $y-z$ cross section, where the number of locations matches the number of unknowns, $C_{ij}$. The resulting square linear system is then solved for $C_{ij}(x_k,t_l)$, corresponding to $y-z$ section at $x=x_k$ and time $t=t_l$. We point out that the decoupling between $x,t$ and $y,z$ in the ansatz for $\partial p/\partial y$ enables solving for $v$ at any $y-z$ cross section, and at any time $t$, without the need to solve simultaneously for $v$ along all the coordinates, $(x,y,z,t)$, which saves on resources and make the method faster and more flexible. \\
%\textcolor{red}{INCLUDE THIS IN THE METHOD FOR GETTING v. In order to arrive at an expression for the $y$-velocity component, we assumed that the channel height does not depend on $z$,  i.e. $D=D(x)$, so that the $z$ velocity component is negligible, compared to $v$, which is turn, scales as $\alpha U$. With $w$ out of the picture, we used to simplified continuity equation and the $y$-component of the momentum equation to arrive at an approximate solution for $v$, but assuming a polynomial dependence of $\partial p/\partial y$ in $z$, where the coefficients, $C_{ij}(x,t)$ (that depend on $x$ and $t$) are to be determined. The $y$-component of the momentum equation is then solved for $v$, using a similar approach to the one used for $u$, in terms of the unknown coefficients. The expression for $v$ is then plugged into the continuity equation which yields the equation governing $C_{ij}(x,t)$. The equation governing $C_{ij}(x,t)$ has all the independent variables, $x, y, z$ and $t$, can be approximately solved  for $C_{ij}(x,t)$ in different ways. We chose to solve for 
%$C_{ij}$ at a given $x$ and $t$ by satisfying the equation at uniformly spaced locations in the $y-z$ cross section, where the number of locations matches the number of unknowns, $C_{ij}$. }
%This endows the method with the added efficiency as it save resources by allowing to obtain a solution for $v$ at selection locations or sections. 

%\input{results}
\extrafloats{100}

\section{Results}\label{section:results} 

The accuracy of the model is compared to CFD simulations for the cases listed in Table \ref{Table:Cases}.  For all cases, a Type I channel geometry is considered (Figure \ref{fig:channelgeo}) with
\begin{equation}
D(x) = 1+ \frac{\gamma_0}{2} \left( 1 - \cos\left(4 \pi k \frac{x}{\lambda} \right)\right) 
\end{equation}
where $\lambda \equiv L=4$, $k=2$, and $\gamma_0=0.05$. Note that all quantities are normalized according to section \ref{ss:nondim} with the dimensions of the channel used as $L=1000 \, \mu m$, $\alpha = 0.05$, $D_0= \frac{\alpha}{\lambda}L$ , and $H_0$= 2$D_0$. 
In Table \ref{Table:Cases}, Cases 1 is the reference case with  $\mathtt{Re}=1, \mathtt{St}=1, \alpha = 0.05, $ and  $\gamma=0.05$. Cases 1-3 aim to explore the performance of the proposed model for the range $1 \leq \mathtt{St} \leq 10$.  Case 1,4, and 5 aim to explore the performance of the proposed model for the range $1 \leq \mathtt{Re} \leq 20$. Case 1,7, 8 and 9 aim to explore the performance of the proposed model for the range $0.05 \leq \gamma \leq 1$. Case 1,10, 11 and 12 aim to explore the performance of the proposed model for the range $0.05 \leq \alpha \leq 0.5$. For all cases, except Case 6, the flow is driven by harmonic forcing of the inlet mass flux according to
%The channel is forced according to two types:
%\begin{itemize}
%\item Harmonic mass flux of the form
\begin{equation}
\dot{m}=\rho U \sin(2\pi f t)
\label{harmonicforcing}
\end{equation}
For Case 6, the flow is driven by ramping up the inlet mass flux according to
%\item Ramp mass flux of the form 
\begin{equation}
\dot{m}=\rho U \left( 1 - {\rm e}^{-ft} \right)
\label{rampforcing}
\end{equation}
%\end{itemize}
where the reference velocity is $U=\frac{\mu \, \mathtt{Re}}{\rho H_0}$ with $\mu=0.001003 \, N{\cdot}s/m^2$, $\rho=998.2 \, kg/m^3$, and the forcing frequency is given by $f=\mathtt{St} \frac{U}{H_0}$.

\begin{table}
\begin{center}
\begin{tabular}{ c c c c c c c}
\hline
 case name & forcing type & $\mathtt{Re}$ & $\mathtt{St}$ &$\alpha$ & $\gamma$ \\ 
             &             &           &         &               &                   \\ 
 \hline
 Case 1 & harmonic & 1 & 1 & 0.05 & 0.05  \\  
 Case 2 & harmonic & 1 & 5 & 0.05 & 0.05  \\ 
 Case 3 & harmonic & 1 & 10 & 0.05 & 0.05  \\  
 \hline 
 Case 4 & harmonic & 5 & 1 & 0.05 & 0.05  \\ 
 Case 5 & harmonic & 20 & 1 & 0.05 & 0.05 \\ 
 Case 6 & ramp & 5 & 1 & 0.05 & 0.05 \\
 \hline 
 Case 7 & harmonic & 1 & 1 & 0.05 & 0.25 \\ 
 Case 8 & harmonic & 1 & 1 & 0.05 & 0.5  \\ 
 Case 9 & harmonic & 1 & 1 & 0.05 & 1.0  \\ 
 \hline 
 Case 10 & harmonic & 1 & 1 & 0.1 & 0.05 \\ 
 Case 11 & harmonic & 1 & 1 & 0.3 & 0.05 \\ 
 Case 12 & harmonic & 1 & 1 & 0.5 & 0.05 \\
 \hline
\end{tabular}
\end{center}
\caption{List of studied cases on Type I channel with their corresponding physical and geometrical parameters}
\label{Table:Cases}
\end{table}

%%

%Errors: \\
%\begin{itemize}
%\item Cases 1,15,16,17: L2err XYU and XZU vs gamma 
%\item Cases 1,15,16,17: L2err XYV vs gamma 
%\item Cases 1,18,19,20: L2err XYU and XZU vs alpha
%\item Cases 1,18,19,20: L2err XYV vs alpha
%\item Cases 1-5: express in words the fact that the error is independent of St for a given Re, alpha, and gamma. (U err is about 0.5\% and verror is about 5\%)
%\end{itemize}
%
%Colormaps:
%\begin{itemize}
%\item Cases 1, 5 (bad mesh?), 15, and 19
%\end{itemize}
%
%Mass flow-rate: Keep the existing figure for case 1 and mention that its error is very small in all cases where the model assumptions are applicable. \\
%
%Pressure profiles for cases 1, 5, 15 and 19. Present departure from the linear profile. 

\begin{figure}
\centering
\includegraphics[width=3.25in]{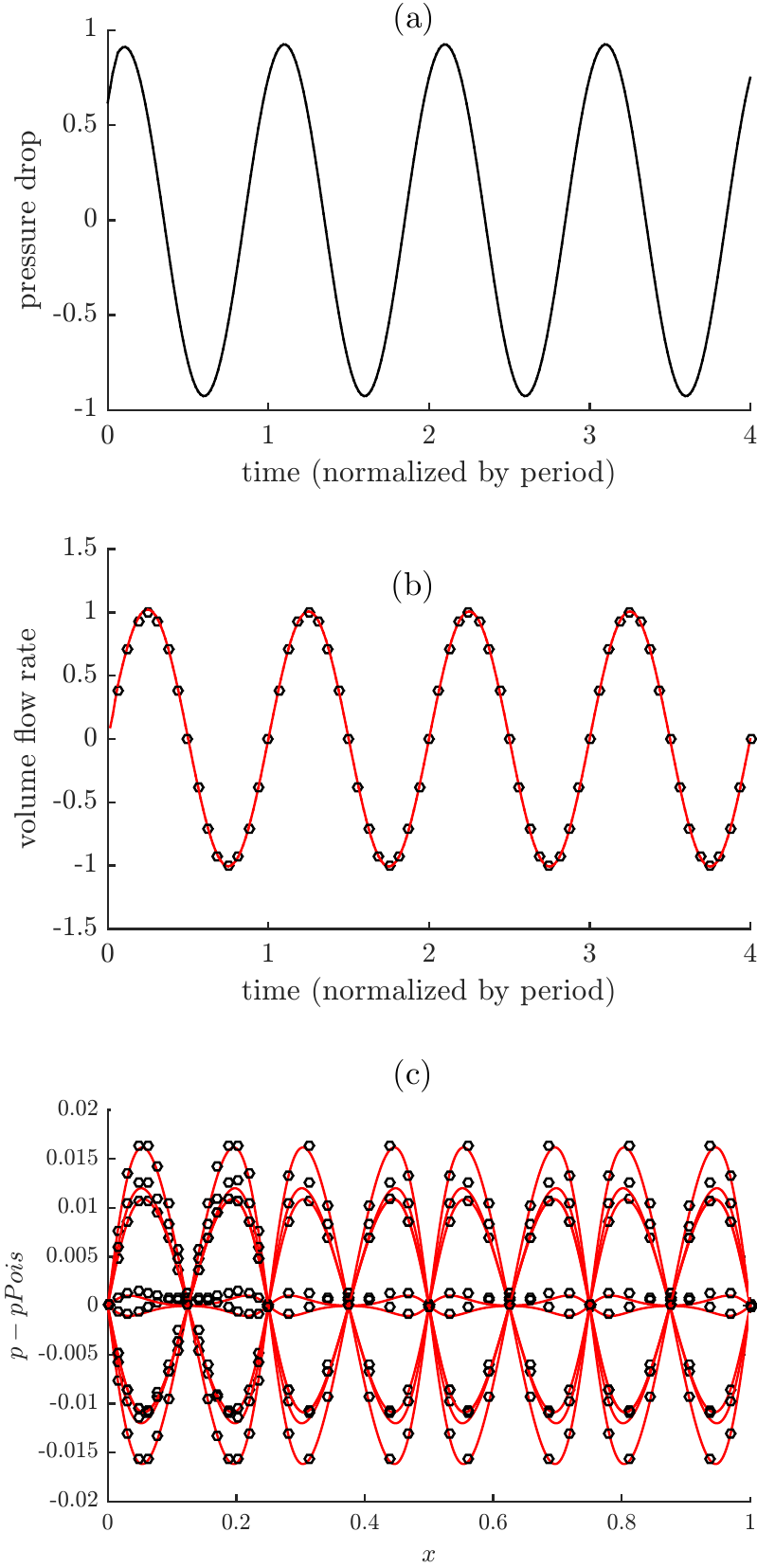}
\caption{(a) Normalized pressure drop history along the channel for Case $9$. (b) Resulting variation in volume flow rate. (c) Variation, over one cycle, along the channel of the departure from the linear profile at $k T/8$, $k=1\rightarrow8$. Solid line: model. Symbols: Ansys.}
\label{fig:Res1}
\end{figure}

%Figure \ref{fig:Res1} illustrates the comparison between the pressure response of the model and the Ansys Fluent simulations for Case 9. 
Figure  \ref{fig:Res1}-(a) plots the time variation of the pressure drop along the channel as predicted by Ansys simulation for the mass flux forcing of Equation (\ref{harmonicforcing}), also plotted using the symbols in Figure  \ref{fig:Res1}-(b). The pressure drop history is then used an input to the reduced order model to predict the unsteady velocity field, the pressure variation along the channel and the mass flow rate. The mass flow rate history predicted by the model matches that of the mass flowrate used to drive the Ansys simulation, as shown in the plot of Figure  \ref{fig:Res1}-(b). In  Figure  \ref{fig:Res1}-(c),  plots of the deviation of the pressure distribution along the channel from the linear (Poiseuille) profile are presented at one eighth intervals of the forcing period, $T=1/f$. The model shows high similarity with the simulation results even at low deviations of order $10^{-3}$.

%The imposed harmonic pressure drop in (a) from \ref{harmonicforcing} results in the oscillatory volume flow rate in (b) where model and simulation match. In (c),  The deviation of the pressure distribution from the linear profile along the channel length is plotted during several time increments of the simulation period. The model also showed high similarity with the simulation results even at low deviations of order $10^{-3}$.

\begin{figure}[!t]
\centering
\includegraphics[width=\textwidth]{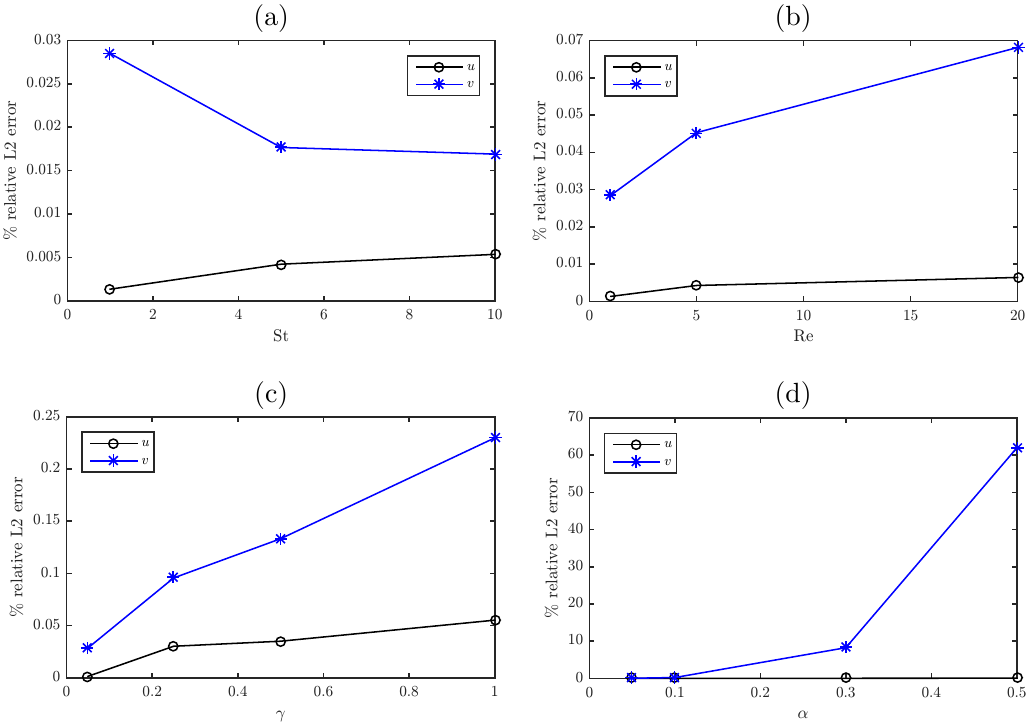}
\caption{Variation of \% relative $L_2$ error in $u$ and $v$ with (a) $\mathtt{St}$ for cases \{1,2,3\}, (b) $\mathtt{Re}$ for cases \{1,4,5\}, (c) $\gamma$ for cases \{1,7,8,9\} and (d) $\alpha$ for cases \{1,10,11,12\}.}
\label{fig:Res2}
\end{figure}

Figure \ref{fig:Res2} shows the variations of the percent relative $L_2$ norm error of the two leading components of the velocity $u$ and $v$ with the dimensional parameters varied in Table \ref{Table:Cases}. The $\%L_2$ errors in $u$ and $v$ are computed respectively as follows 
\begin{align}
\%L_2err(u) &= 100 \times \frac{\sum_{i=1}^{8} \sum_{j=1}^8 \sum_k (u_m(t_i,x_j,y_k,z_k) - u_n(t_i,x_j,y_k,z_k))^2}{\sum_{i=1}^{8} \sum_{j=1}^8 \sum_k u_n(t_i,x_j,y_k,z_k)^2}
\label{eq:L2Erru} \\
\%L_2err(v) &= 100 \times \frac{\sum_{i=1}^{8} \sum_{j=1}^8 \sum_k (v_m(t_i,x_j,y_k,z_k) - v_n(t_i,x_j,y_k,z_k))^2}{\sum_{i=1}^{8} \sum_{j=1}^8 \sum_k v_n(t_i,x_j,y_k,z_k)^2}
\label{eq:L2Errv}
\end{align}
The error is evaluated as the sum of square differences between the  model solution (subscript $m$) and Ansys Fluent results (subscript $n$) computed at $y-z$ cross sections that are evenly spaced over the wavelength ($x_{j+1}-x_j= \lambda/8$) and over evenly spaced intervals of the forcing period ($t_{i+1}-t_i = T/8$).
Figure \ref{fig:Res2}-(a) shows that the model is accurate over the considered range of $\mathtt{St}$ in Cases 1-3, where the $\%L_2$ error is less than $0.005\%$ for $u$ and less than $0.03\%$ for $v$. This behavior is expected since the accuracy of the model  depends on $\alpha$ and $\gamma$, both of which assume the small value of 0.05 for Cases 1-3.

\begin{figure}
\centering
\includegraphics[width=0.84\textwidth]{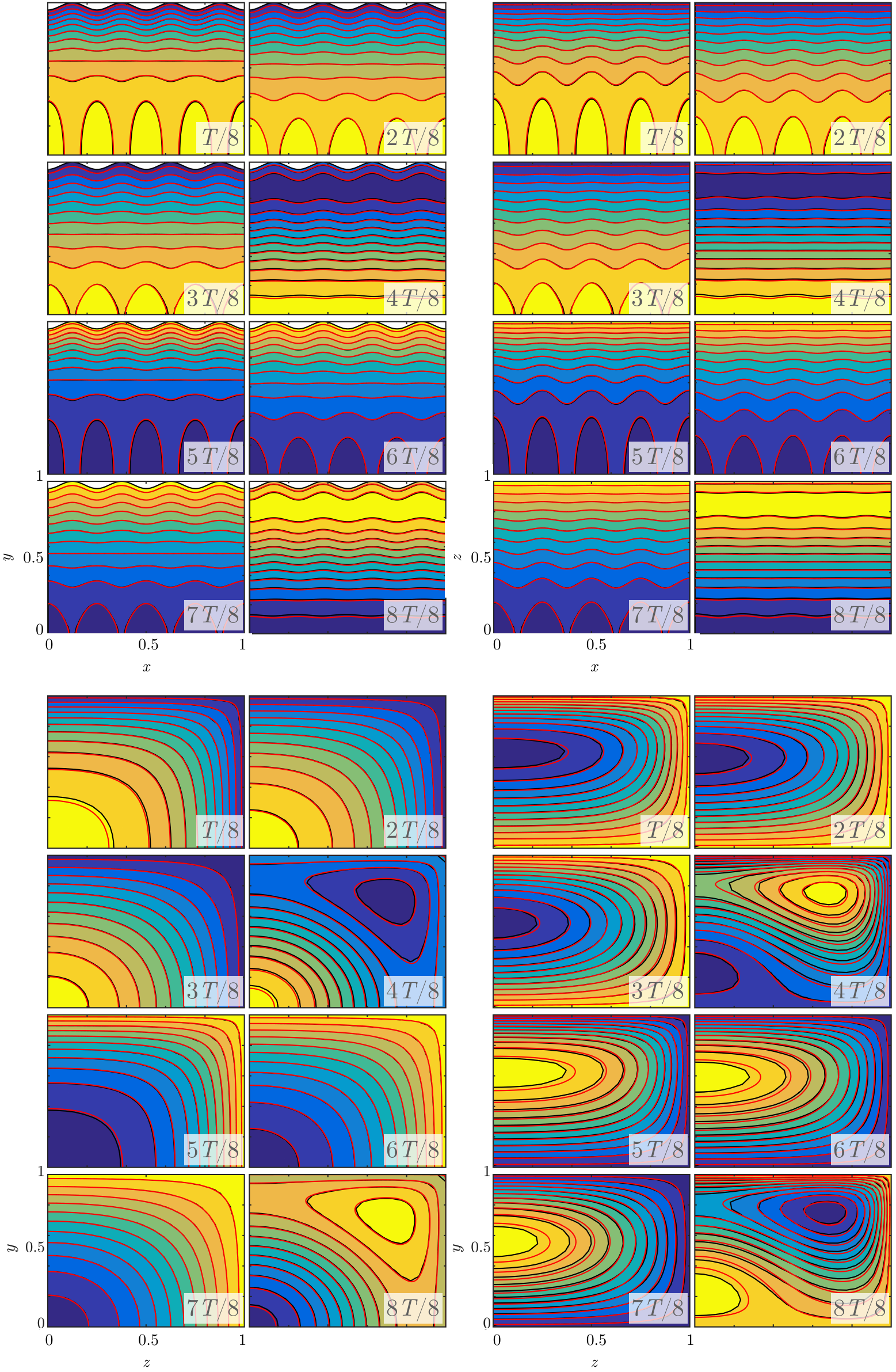}
\caption{Contours of $u_0$ and $v_0$ at $T/8$ time intervals of one cycle for Case 1.  Top-left: $u_0$, $xy$-plane of symmetry ($z=0$),  Top-right: $u_0$, $xz$-plane of symmetry ($y=0$), Bottom-left: $u_0$, $yz$-plane at $x=1.875$, Bottom-right: $v_0$,  $yz$-plane at $x=1.875$. }
\label{fig:Case1}
\end{figure}

\begin{figure}
\centering
\includegraphics[width=0.83\textwidth]{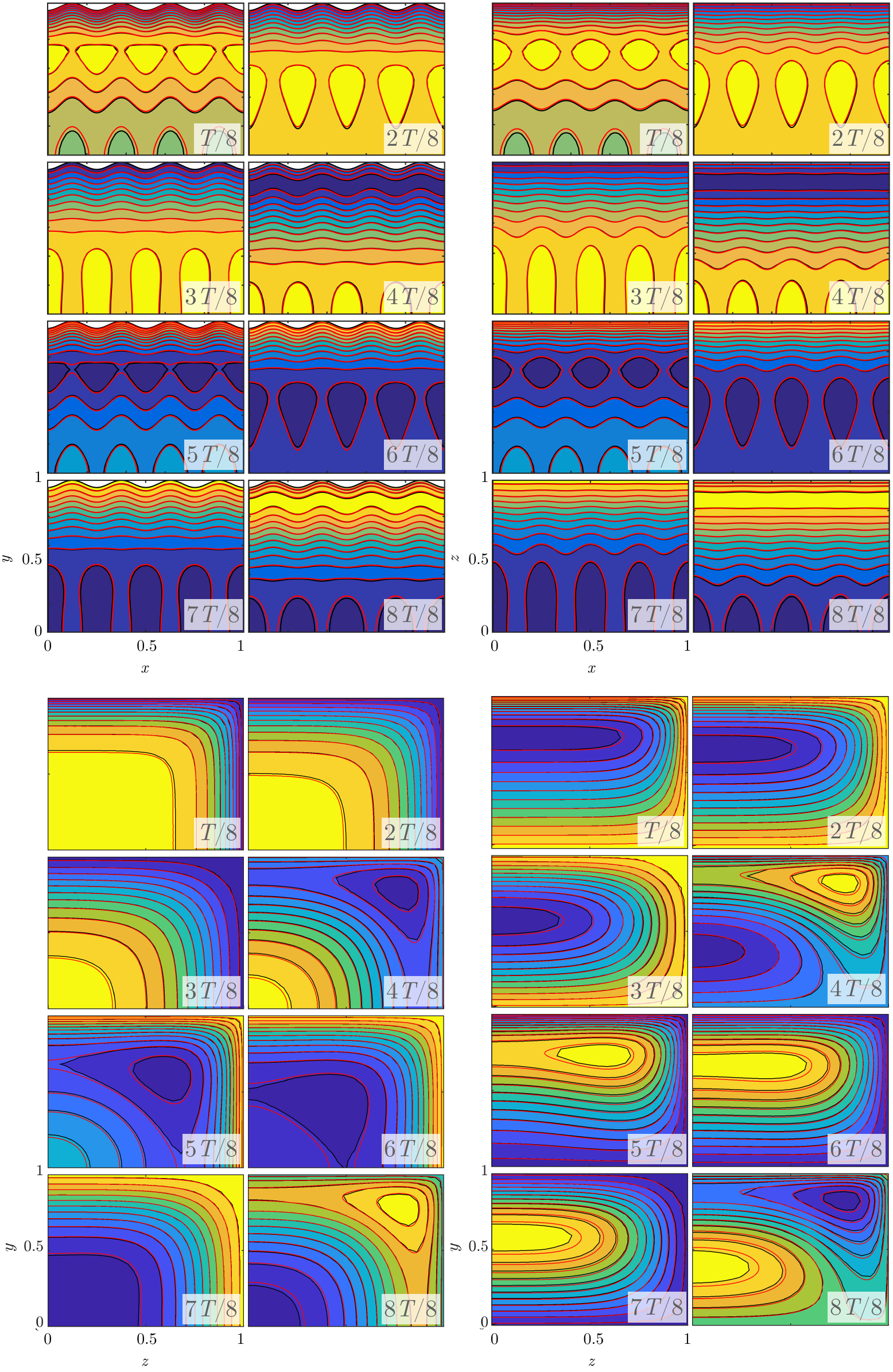}
\caption{Contours of $u_0$ and $v_0$ at $T/8$ time intervals of one cycle for Case 2.  Top-left: $u_0$, $xy$-plane of symmetry ($z=0$),  Top-right: $u_0$, $xz$-plane of symmetry ($y=0$), Bottom-left: $u_0$, $yz$-plane at $x=1.875$, Bottom-right: $v_0$,  $yz$-plane at $x=1.875$. }
\label{fig:Case2}
\end{figure}

\begin{figure}
\centering
\includegraphics[width=0.84\textwidth]{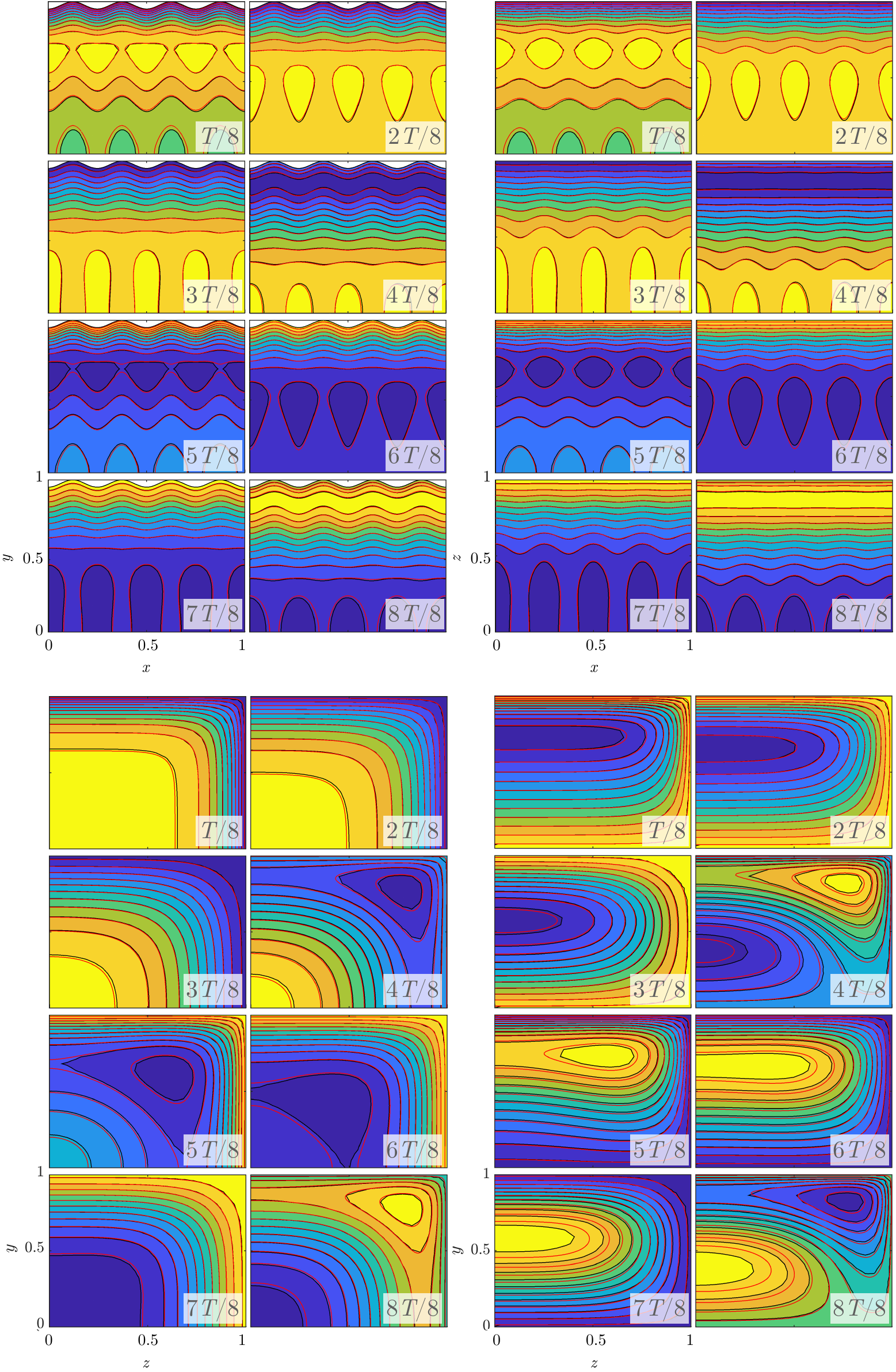}
\caption{Contours of $u_0$ and $v_0$ at $T/8$ time intervals of one cycle for Case 4.  Top-left: $u_0$, $xy$-plane of symmetry ($z=0$),  Top-right: $u_0$, $xz$-plane of symmetry ($y=0$), Bottom-left: $u_0$, $yz$-plane at $x=1.875$, Bottom-right: $v_0$,  $yz$-plane at $x=1.875$. }
\label{fig:Case4}
\end{figure}

Figure \ref{fig:Res2}-(b) presents the variation of the  $\%L_2$ errors with $\mathtt{Re}$ for Cases 1, 4 and 5. The error in $u$ remains less than $0.005\%$ while the error in $v$ is less than $0.045\%$. We notice that the difference between the relative errors of the axial and transverse components of velocity mimics the difference in their scales as the error in $v$ is about one order of magnitude greater than that of $u$. Note that the $L_2$ error in $v$ is normalized by the $L_2$ norm of $v$ which scales as $\alpha U$. The modest dependence of $\%L_2$ error on $\mathtt{Re}$ and $\mathtt{St}$ numbers is clearly visualized in Figures \ref{fig:Case1}, \ref{fig:Case2}, and \ref{fig:Case4}. The model solution contours, represented in red color, show pronounced agreement with the simulation results, with a subtle decline of accuracy for the transverse component of the velocity.  \\
 
\begin{figure}
\centering
\includegraphics[width=0.83\textwidth]{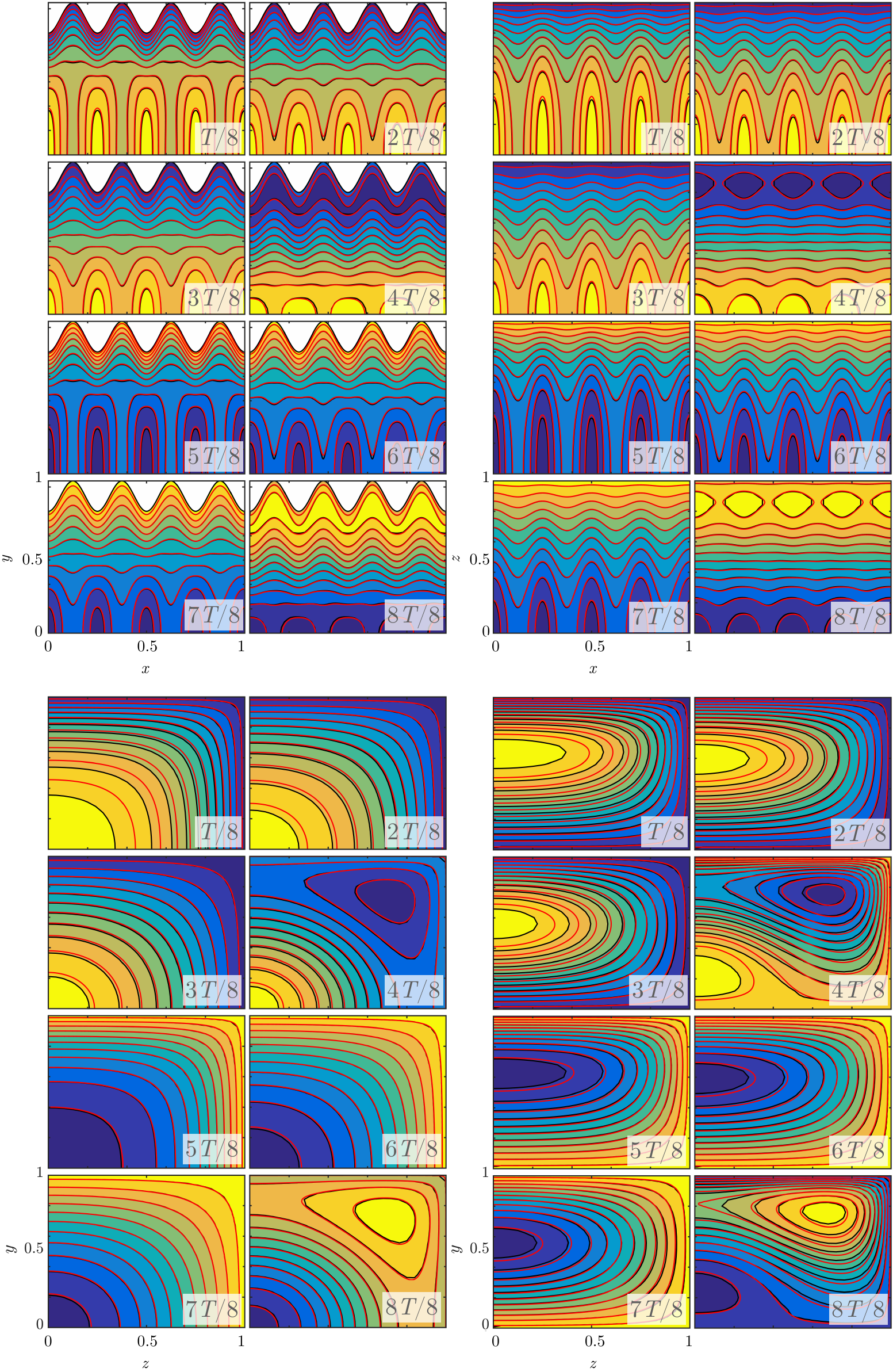}
\caption{Contours of $u_0$ and $v_0$ at $T/8$ time intervals of one cycle for Case $7$.  Top-left: $u_0$, $xy$-plane of symmetry ($z=0$),  Top-right: $u_0$, $xz$-plane of symmetry ($y=0$), Bottom-left: $u_0$, $yz$-plane at $x=3.125$, Bottom-right: $v_0$,  $yz$-plane at $x=3.125$. }
\label{fig:Case7}
\end{figure}

\begin{figure}
\centering
\includegraphics[width=0.84\textwidth]{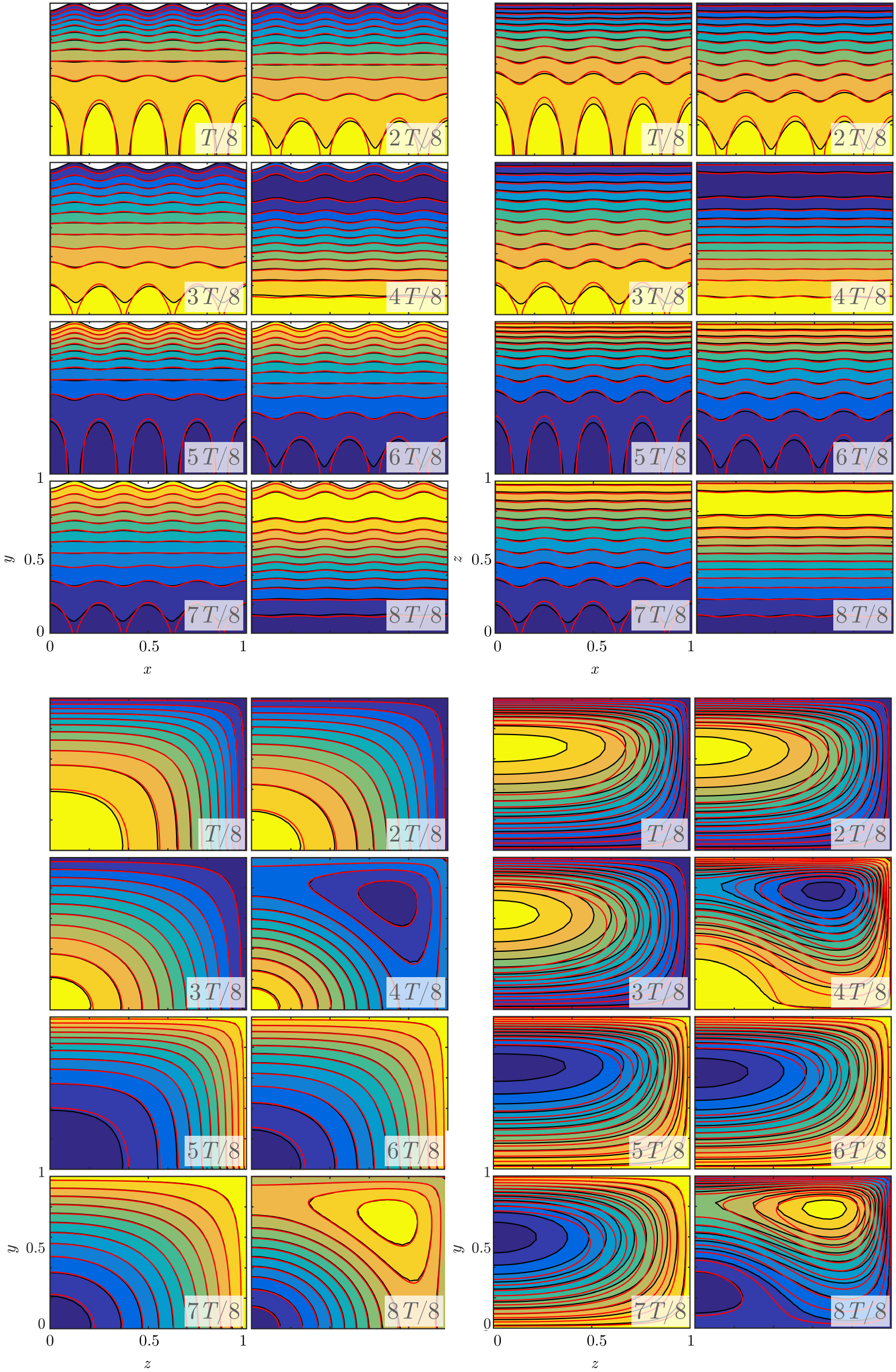}
\caption{Contours of $u_0$ and $v_0$ at $T/8$ time intervals of one cycle for Case 11.  Top-left: $u_0$, $xy$-plane of symmetry ($z=0$),  Top-right: $u_0$, $xz$-plane of symmetry ($y=0$), Bottom-left: $u_0$, $yz$-plane at $x=3.125$, Bottom-right: $v_0$,  $yz$-plane at $x=3.125$. }
\label{fig:Case11}
\end{figure}

On the other hand, Figures \ref{fig:Res2}-(c) and (d)  show a stronger dependence of the $\%L_2$ error on the dimensionless geometric parameters $\gamma$ and $\alpha$. For the range $0.05 \leq \gamma \leq 1$ in Cases 1, 7, 8, and 9 of Table \ref{Table:Cases}, the $\%L_2$ error, plotted in Figure-(c), increases fro  $0.002\%$ to $0.05\%$ for the axial velocity and from $0.03\%$ to $0.24\%$ for the transverse velocity.  For Case 7, corresponding to $\mathtt{Re}=1, \mathtt{St}=1, \alpha = 0.05, $ and  $\gamma=0.25$, This accuracy of the model can also be assessed by inspecting Figure \ref{fig:Case7}, which shows the model-predicted contours (solid lines) on top of the Ansys-predicted color distributions of $u$ and $v$ at selected $x-y$, $y-z$ and $x-z$ planes at $T/8$ time intervals of one cycle.\\
For the range $0.05 \leq \alpha \leq 0.5$ in Cases 1, 10, 11 and 12 of Table \ref{Table:Cases}, the $\%L_2$ error in $v$, plotted in Figure-(d), increases abruptly beyond $\alpha=0.1$ to reach $60\%$ for $\alpha=0.5$. Presenting the error in this way exaggerates the discrepancy. The error in $v$ increases to reach 60\% because the average $v$ decreases for larger $\alpha$, which means that the denominator in Equation (\ref{eq:L2Errv}) decreases as $\alpha$ increases while keeping $\mathtt{Re}, \mathtt{St}$ and $\gamma$ constant. Actually for a fully developed inertia-free flow, and for small values of $\gamma$, the streamlines are nearly parallel to the channel axis for $y/D(x)<<1$, and as $y/D(x)$ approaches 1, the streamlines become nearly parallel to the channel wall. As $\alpha$ increases for a given small $\gamma$, the effect of the curvature of the wall on the streamlines gets smaller. For $\alpha/\gamma>>1$, most of the streamlines are parallel to the channel axis, which implies that $v \simeq 0$ in most of the domain, except near the wall. As such, a smaller error in $v$ gets significantly amplified when normalized by the $L_2$ norm of $v$. Upon inspecting Figure \ref{fig:Case11}, one can observe that the model-predicted contours (solid lines) of $u$ are in good agreement with the color distributions predicted by Ansys-Fluent simulations at selected $x-y$, $y-z$ and $x-z$ planes at $T/8$ time intervals of one cycle. This is not the case for contours of $v$, where the model-predicted contours diverge considerably from the Ansys solution, as shown in the bottom right subfigure. A pointed out in the discussion above, the values of $v$ in this cases are very small. 

%\begin{figure}[t]
%\includegraphics[width=\textwidth]{figures/u0_Zplane_y0_lr}
%\label{u0contourY0}
%\caption{CONTOUR OF $u_0$ VELOCITY AT $xy$-PLANE OF Y=0 AT CONSECUTIVE TIMESTEPS}
%\end{figure.}

%\begin{figure}[t]
%\includegraphics[width=\textwidth]{figures/u0_XYplane_z12p5_lr}
%\label{u0contourY0}
%\caption{CONTOUR OF $u_0$ VELOCITY AT $xz$-PLANE OF Z=12.5 AT CONSECUTIVE TIMESTEPS}
%\end{figure}

%The following Figures are added by Ali
%%%%%%%%%%%%%%%%%%%%%%%%%%%%%%%%%%%%%%%%%%%%%%%%%%%%%%%%%%%%%%%%%%%%%%%
%%%%%%%%%%%%%%% HARMONIC FIGURES %%%%%%%%%%%%%%%%%%%%%%%%%%%%%%%%%%%%%

\begin{figure}
\centering
\includegraphics[width=3.25in]{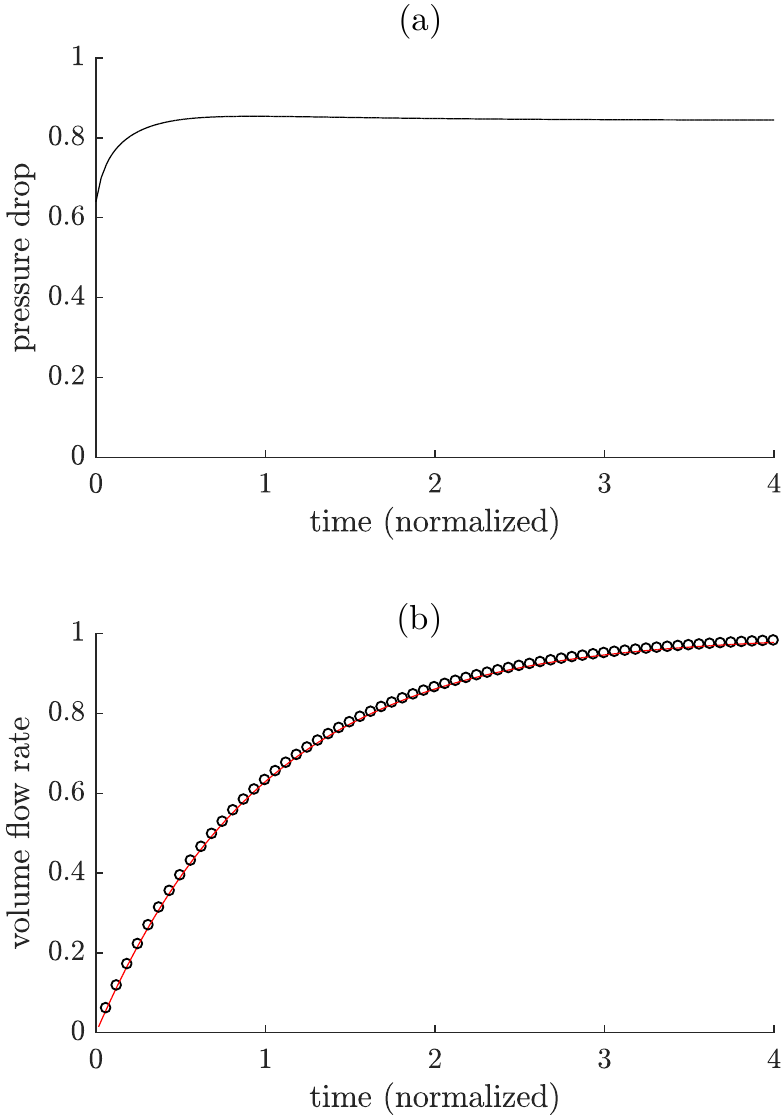}
\caption{(a) Normalized pressure drop history along the channel for Case 6. (b) Resulting variation in volume flow rate. Solid line: model. Symbols: Ansys.}
\label{fig:PCase6}
\end{figure}

\begin{figure}
\centering
\includegraphics[width=0.84\textwidth]{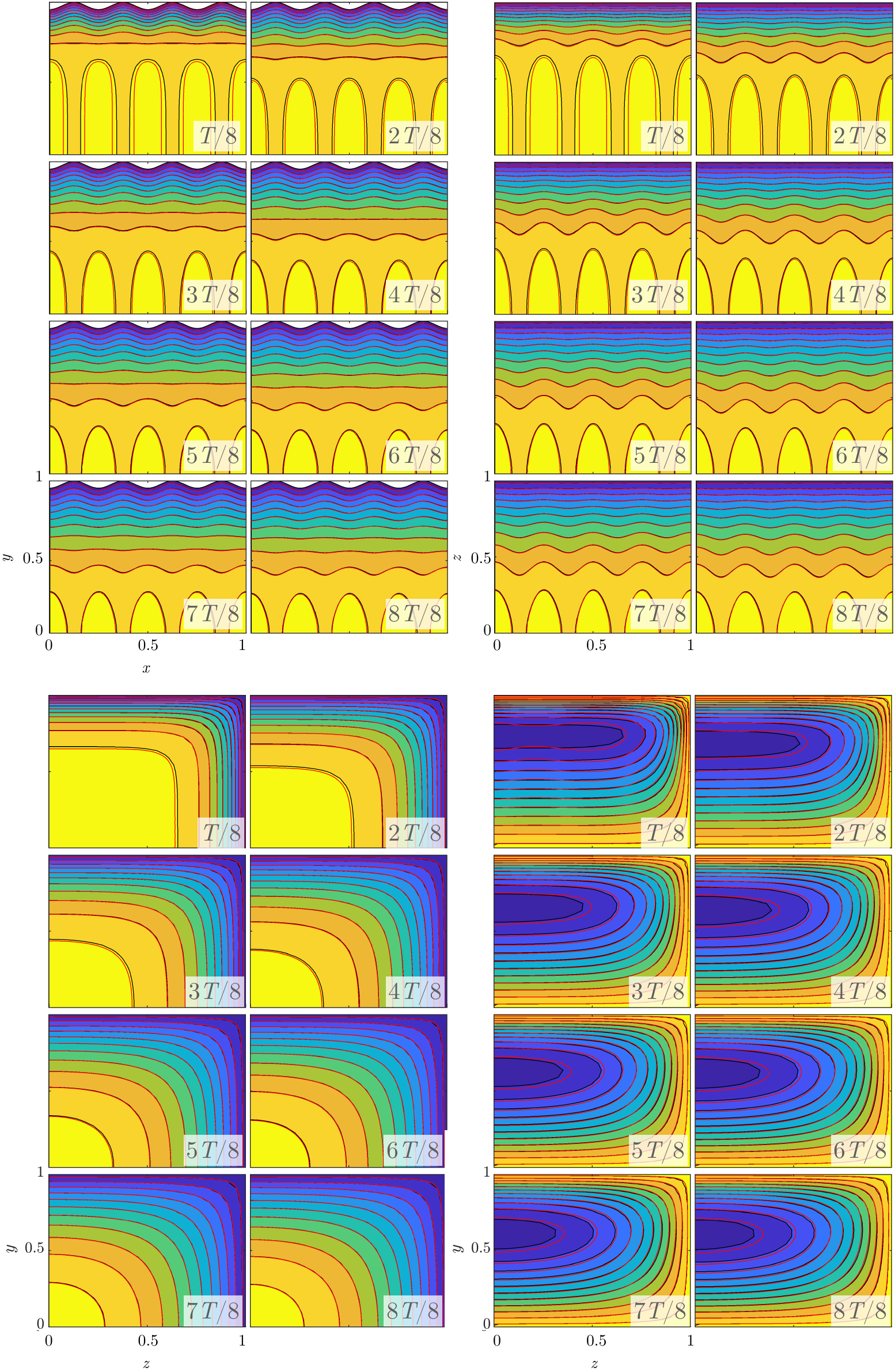}
\caption{Contours of $u_0$ and $v_0$ at $T/8$ time intervals of one cycle for Case 6.  Top-left: $u_0$, $xy$-plane of symmetry ($z=0$),  Top-right: $u_0$, $xz$-plane of symmetry ($y=0$), Bottom-left: $u_0$, $yz$-plane at $x=1.75$, Bottom-right: $v_0$,  $yz$-plane at $x=1.75$. }
\label{fig:Case6}
\end{figure}

For Case 6 of Table \ref{Table:Cases}, the flow is driven by an inlet mass flux obeying equation (\ref{rampforcing}). The corresponding time variation of the pressure drop along the channel as predicted by Ansys simulation experiences a sharp rise to it steady value, as plotted in Figure \ref{fig:PCase6}-(a). The pressure drop history is then used an input to the reduced order model to predict the unsteady velocity field, the pressure variation along the channel and the mass flow rate. The mass flow rate history predicted by the model matches that of the mass flowrate used to drive the Ansys simulation, as shown in the plot of Figure \ref{fig:PCase6}-(b). The accuracy of the model is also apparent in Figure \ref{fig:Case6}, where the model-predicted contours of $u$ and $v$ match those predicted by the Ansys simulation with a slight deviation on the transverse velocity component $v$.   

%%%%%%%%%%%%%%%%%%%%%%%%%%%%%%%%%%%%%%%%%%%%%%%%%%%
%\section{Discussion}

\section{Conclusion}\label{section:Conclusion}
In this paper, we presented a simplified model for predicting the dynamic behavior of low $\alpha {\mathtt Re}$ flows in three-dimensional channels of slowly varying cross section. The model is based on an asymptotic expansion of the governing equations in the aspect ratio of the channel $\alpha << 1$. 

The solution of these leading order equations is a hybrid analytical-numerical solution. 
The leading order axial velocity $u$ is obtained using a finite Fourier transform in the plane orthogonal to the flow direction on the momentum equation; the pressure is obtained via a recursive numerical scheme that uses the mass equation and the expression of the solution to the axial velocity. 
In order to solve the $y$-component of the momentum equation for $v$, we use and ansatz for $\partial p/\partial y$ in the form of a polynomial in $y, z$, where the coefficients are dependent on $x$ and $t$. We then use the continuity equation while assuming that $w \simeq 0$, which is consistent with our assumption that the channel height only depends on $x.$ This allows us to solve for the unknown coefficients. 
In this fashion, we obtain $v$ at any $y-z$ cross-section, at any instant, without having to solve for the entire $v(x,y,z,t),$ thus saving on resources and speed.  

In contrast to previous works, the proposed approach allows for a general time dependence of the driving pressure force, provided that the advective inertia term remains much smaller than the driving force. As such, the model accurately predicts the dynamic flow behavior in response to fast temporal changes in  $\Delta p$, which are characterized by large values of the Strouhal number, associated with the high frequency components comprising the pressure drop.  

The accuracy of the model is assessed by inspecting the behavior of the relative $L_2$ norm of the error in the $u$ and $v$ components, where the true solution is obtained from converged Ansys Fluent simulations. In addition, contours of the velocity components at selected cross sections, resulting from the simplified model are compared with those of Ansys. The results demonstrate that the proposed model is capable of accurately predicting the velocity field for most cases considered. In fact, in the ranges  (${\mathtt St} <10$, ${\mathtt Re} < 20$, $
\gamma < 1$, $\alpha<0.5$), the relative percent $L_2$ error in $u$ (and $p$) is less than $1\%$. 
As for the error in $v$, the relative percent $L_2$ error in $v$ is less than $10\%$ for all the cases considered in the ranges (${\mathtt St} <10$, ${\mathtt Re} < 20$, $
\gamma < 1$, $\alpha<0.3$), except when $\alpha=0.5$ and for small $\gamma$.
  As discussed above, the relative error in $v$ increases with $\alpha$, for small $
\gamma$, and this is because $v$ itself is very small for this case. In fact, the streamlines are nearly parallel to the $x$-axis over most of the domain. We note that the way we presented the error dependence in Figure \ref{fig:Res2} can also be used to gain information about the error sensitivity to variations in $\mathtt{Re}$, $\mathtt{St}$, $\alpha$, $\gamma$.\\
 The speed of the simplified model is around two orders of magnitude larger than that of the detailed 3D CFD simulations of Ansys Fluent.

 % because $v$ is close to 0 in most of the domain. except for the $v$ velocity component for large $\alpha$, where an error of around 60\% is incurred for $\alpha=0.5$. 

We envision that future directions for this work can proceed along two tracks: a methodology track and an applications track. (i) In the methodology track, directions that are worth exploring include the possibility of relaxing the condition on $w$ by allowing for variations of the channel cross section along $z$. Another direction is to further investigate whether we can use a better alternative to the low degree polynomial ansatz for $\partial p/\partial y$, and explore alternative ways to solve for the unknown polynomial coefficients. %conversion of the equation governing the unknown coefficient to a linear system by choosing to satisfy the equation at uniformly spaced points, where the number of these points matches the number of unknowns. 
(ii) in the application track, a potentially impactful future direction is to couple this simplified model to an advection diffusion solver. The solver can then be embedded within a framework to quickly assess proposed mixing/heat transfer mechanisms in micro-fluidic applications and arrive at conditions that optimize the efficiency.

%employ the velocity field obtained using this simplified approach in another simplified model for solving the advection-diffusion equation. The two simplified models can then be embedded within a framework to quickly assess proposed mixing/heat transfer mechanisms in micro-fluidic applications and arrive at conditions that optimize the efficiency. 

%talk about future directions and applications
%Things that can go there: (1) difficulty of obtaining the third component without a scaling assumption. (2) a clear idea of how to use this algorithm for optimizing a mixing coefficient? without doing it but promise to do it as an extension.
%Should we combine the discussion with the conclusion here?

\section{Acknowledgments}
This work was supported by LAU SRDC grant SRDC-r-2017-21 \\

\newpage

 \bibliographystyle{elsarticle-num} 

% Here's where you specify the bibliography database file.
% The full file name of the bibliography database for this
% article is asme2e.bib. The name for your database is up
\bibliography{THEbib}%% else use the following coding to input the bibitems directly in the

\end{document}